\documentclass[conference]{IEEEtran}
\IEEEoverridecommandlockouts
\usepackage{cite}
\usepackage{amsmath,amssymb,amsfonts}
\usepackage{algorithmic}
\usepackage{graphicx}
\usepackage{textcomp}
\usepackage{xcolor}
\usepackage[ruled,linesnumbered]{algorithm2e}
\usepackage{float}
\usepackage{subcaption}
\usepackage{hyperref}
\newcommand{\zqy}[1]{\textbf{\color{red}{#1}}}
\def\BibTeX{{\rm B\kern-.05em{\sc i\kern-.025em b}\kern-.08em
    T\kern-.1667em\lower.7ex\hbox{E}\kern-.125emX}}

\IEEEoverridecommandlockouts\IEEEpubid{\makebox[\columnwidth]{ 979-8-3503-0322-3/23/\$31.00 $\copyright$2023 IEEE \hfill}\hspace{\columnsep}\makebox[\columnwidth]{ }}
\begin{document}

\title{Energy and Time-Aware Inference Offloading for DNN-based Applications in LEO Satellites}
% {\footnotesize \textsuperscript{*}Note: Sub-titles are not captured in Xplore and
% should not be used}
% \thanks{Identify applicable funding agency here. If none, delete this.}
% }
% Identify applicable funding agency here. If none, delete this.

\author{\IEEEauthorblockN{Yijie Chen, Qiyang Zhang, Yiran Zhang, Xiao Ma, Ao Zhou\\
\textit{State Key Laboratory of Networking and Switching Technology}\\
\textit{Beijing University of Posts and Telecommunicatons,Beijing, China}\\
\{yijiechen;qyzhang;sgwang\}@bupt.edu.cn, \{yijiechen;qyzhang;sgwang\}
}

% \IEEEauthorblockA{\textit{dept. name of organization (of Aff.)} \\
% % \textit{name of organization (of Aff.)}\\
% City, Country \\
% email address or ORCID}
% \and
% \IEEEauthorblockN{2\textsuperscript{nd} Given Name Surname}
% \IEEEauthorblockA{\textit{dept. name of organization (of Aff.)} \\
% \textit{name of organization (of Aff.)}\\
% City, Country \\
% email address or ORCID}
% \and
% \IEEEauthorblockN{3\textsuperscript{rd} Given Name Surname}
% \IEEEauthorblockA{\textit{dept. name of organization (of Aff.)} \\
% \textit{name of organization (of Aff.)}\\
% City, Country \\
% email address or ORCID}
% \and
% \IEEEauthorblockN{4\textsuperscript{th} Given Name Surname}
% \IEEEauthorblockA{\textit{dept. name of organization (of Aff.)} \\
% \textit{name of organization (of Aff.)}\\
% City, Country \\
% email address or ORCID}
% \and
% \IEEEauthorblockN{5\textsuperscript{th} Given Name Surname}
% \IEEEauthorblockA{\textit{dept. name of organization (of Aff.)} \\
% \textit{name of organization (of Aff.)}\\
% City, Country \\
% email address or ORCID}
% \and
% \IEEEauthorblockN{3\textsuperscript{rd} Given Name Surname}
% \IEEEauthorblockA{\textit{dept. name of organization (of Aff.)} \\
% \textit{name of organization (of Aff.)}\\
% City, Country \\
% email address or ORCID}
}

\maketitle

\begin{abstract}
%Recently, Low Earth Orbit (LEO) satellites have experienced rapid development and widespread application, leading to Deep Neural Network (DNN) models emerging as the dominant technology for remote sensing satellite image recognition, particularly in real-time target detection supporting LEO satellites. 
In recent years, Low Earth Orbit (LEO) satellites have witnessed rapid development, with inference based on Deep Neural Network (DNN) models emerging as the prevailing technology for remote sensing satellite image recognition.
%However, the substantial computational energy and time demands of DNN models pose significant challenges: burdening satellites with limited power intake and impeding the timely completion of real-time tasks. 
However, the substantial computation capability and energy demands of DNN models, coupled with the instability of the satellite-ground link, pose significant challenges, burdening satellites with limited power intake and hindering the timely completion of tasks.
Existing approaches, such as transmitting all images to the ground for processing or executing DNN models on the satellite, is unable to effectively address this issue.
%Considering the distinctive characteristics of various tasks and dynamic connection of satellite-ground,
By exploiting the internal hierarchical structure of DNNs and treating each layer as an independent subtask, we propose a satellite-ground collaborative computation partial offloading approach to address this challenge.
We formulate the problem of minimizing the inference task execution time and onboard energy consumption through offloading as an integer linear programming (ILP) model. 
The complexity in solving the problem arises from the combinatorial explosion in the discrete solution space.
To address this, we have designed an improved optimization algorithm based on branch and bound.
Simulation results illustrate that, compared to the existing approaches, our algorithm improve the performance by 10\%-18\%.\\
% With the increasing adoption of Low Earth Orbit (LEO) satellites, the challenges associated with computation and task offloading on these satellites have become prominent. This study aims to address the issues related to limited computing resources and constrained energy acquisition on LEO satellites. In this paper, a comprehensive model encompassing both latency and energy consumption is introduced to tackle these challenges, followed by its formulation as an integer linear programming problem. This paper propose an enhanced algorithm for integer linear programming utilizing branch and bound technique (ILPB) to effectively solve this problem. Through extensive simulations, the results show the substantial reduction in latency and satellite energy consumption achieved by the proposed ILPB algorithm.\\
\end{abstract}

 
\section{Introduction}

With the continuous technological advancements and the growing demand for space exploration, Low Earth orbit (LEO) satellites, exemplified by constellations like SpaceX, OneWeb, and Telesat \cite{b1}, are rapidly evolving.
LEO satellites serve various purposes with Earth observation being a typical one. This involves capturing ground images and using a Deep Neural Network (DNN) model to infer occurrences such as forest fires, terrain, and geomorphic changes \cite{b17,b18}.

The prevailing approach involves transmitting all images to the ground for processing, utilizing a traditional method known as the “bent-pipe” architecture \cite{b3}. 
However, the capacity of the satellite-ground link is limited, and the connection between satellites and the ground links is often unreliable and periodic, resulting in a significant amount of data remaining that cannot be timely transmitted.
An alternative solution involves executing DNN models on the satellite \cite{b7}. 
However, the majority of LEO satellites are equipped with limited computing power due to their small size and reliance on solar energy collection, making it extremely challenging to run the entire large-scale DNN models on energy-limited satellites \cite{b19}. 
As an alternative, small-scale DNN models are considered, but they may not guarantee the same level of inference accuracy  \cite{b20}.

To tackle this issue, we exploit the hierarchical structure of DNNs and treat each layer as an independent subtask.
The input matrices of one layer are derived from the output matrices of the preceding layer. 
A feasible strategy involves executing certain layers on the LEO satellite while offloading the remaining layers and the intermediate output to the ground.
Considering that the input matrices of most layers in the DNN model tend to decrease compared to the initial input, this strategy effectively utilizes the limited resources of LEO satellites to significantly reduce the transmitted data size.

However, accurately selecting the appropriate layers for offloading is a challenging task due to the following factors: 1) notable variations exist in the internal structure among different DNNs; 2) 
due to the diverse nature of DNN layers, different offloading strategies yield diverse performances;
3) the constrained computation capability, link capacity, and energy of LEO satellites all contributes to the execution time and onboard energy consumption, making it a coupled and complex problem.
To overcome these challenges, we establish a general %一般都是普通的都是，general
model to ensure more generalizable results and propose an energy and time-aware inference offloading algorithm to address this problem.

The main contributions of this paper are as follows.

\begin{itemize}
\item 
To the best of our knowledge, we are the first to investigate inference offloading based on the layered architecture of the DNN model in LEO satellites. 
%This model improves upon the shortcomings of existing solutions and leverages hierarchical offloading of DNN models.
% We examine the challenges associated with current satellite-ground collaborative inference tasks offloading, and propose a task model that leverages hierarchical offloading of DNNs based on the identified shortcomings of these existing solutions.
% \item We examine the challenges associated with current satellite-ground collaborative inference tasks offloading and provide an overview of the problems encountered in existing solutions. 
% % These challenges include inadequate task timeliness, excessive utilization of satellite resources, and the requirement for separate and complete inference processes for both on-board and ground-based processing.
% \item We propose a task model that leverages hierarchical offloading of DNNs based on the identified shortcomings of existing solutions. This model incorporates a comprehensive assessment of both latency and energy consumption variables to evaluate the overall effectiveness of the outcomes.
\item  We formulate the problem as a constrained integer linear programming (ILP) problem and propose an improved algorithm based on branch and bound to address it.
\item We evaluate the performance of the proposed algorithm through simulations, demonstrating promising experimental results that can support energy-efficient and time-saving offloading.
\end{itemize}

\section{Background and Motivation}
\textbf{“bent pipe” architecture.} 
Most current LEO satellites still employ a communication model known as “bent pipe” architecture. 
In this architecture, satellites serve as data-transmitting relays without data processing capabilities. 
In particular, Cheng et al. \cite{b4} proposed a satellite-ground integrated network edge computing architecture that utilizes satellites to enable access to cloud computing resources. 
Giuliari et al. \cite{b5} introduced a kind of inter-networking approach, in which ground stations work as access points or gateways for satellite networks, in order to improve the efficiency of satellite data transmission. 

Nonetheless, the data transmission rate from the satellite to the ground is significantly hindered by an unstable communication link and intermittent availability, resulting in a much lower transmission rate compared to the data generation rate on the satellite.
Consequently, this led to a substantial increase in the overall latency of entire tasks.

\textbf{Satellite edge computing.}
Satellite edge computing has been developed to facilitate on-orbit processing.
Xu et al. \cite{b6} presented a space-ground-sea integrated network architecture, utilizing satellites as computational nodes for relevant tasks, with the aim of minimizing user execution latency.
Denby et al. \cite{b7} proposed an Orbital edge computing system that performs partial processing and data filtering on satellites. This system aims to enhance downlink connections' efficiency in saturated satellites while considering energy limitations and payload impacts. 
In the existing satellite edge computing schemes, the entire inference process is executed on the satellite. 
This approach imposes a heavy burden on the satellite due to its limited computing resources and low energy acquisition capabilities.

To address the above issues, 
We consider the varying computational requirements and data volume at different layers of DNNs and investigate a computation offloading strategy aimed at reducing energy consumption and latency.
%This approach effectively mitigates the issue of limited energy resources on satellites and restricted satellite-to-ground transmission, while minimizing energy consumption on valuable satellite resources and task latency.
\begin{figure}[t]
% \centerline{\includegraphics{fig.png}}
\vspace*{-\abovecaptionskip}
\centering
\includegraphics[width=0.5\textwidth]{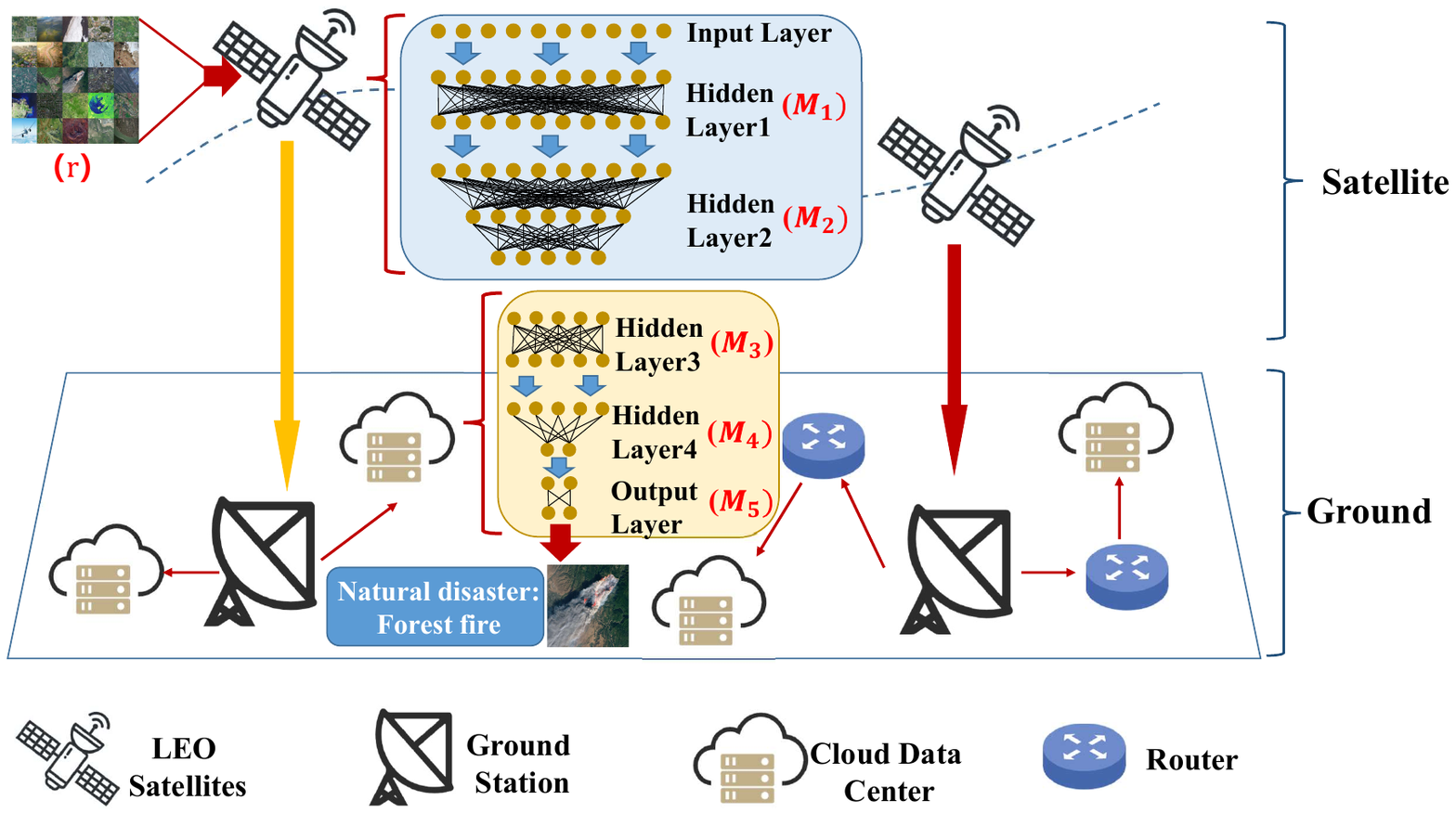}
\caption{Architecture of Satellite-Ground Collaborative Computing.
}
\label{fig}
\end{figure}
\section{SYSTEM MODEL AND PROBLEM DEFINITION}\label{sec:modeling}
In this section, we first introduce the system model, context, and notation used in this work. We then provide a more detailed description of the problem being considered.
% Before you begin to format your paper, first write and save the content as a 
% separate text file. Complete all content and organizational editing before 
% formatting. Please note sections \ref{AA}--\ref{SCM} below for more information on 
% proofreading, spelling and grammar.

% Keep your text and graphic files separate until after the text has been 
% formatted and styled. Do not number text heads---{\LaTeX} will do that 
% for you.

\subsection{System Model}\label{AA} 
Our satellite-ground collaborative architecture consists of LEO satellites, ground stations, and cloud data centers, as shown in Fig.~\ref{fig}.
Every LEO satellite is equipped with computationally and storage-capable payloads. 
This enables edge-like satellites to process partial images and serve as data transmission routers.  
When a satellite comes within the connected range of a ground station, it becomes capable of transmitting data to the ground station.
Cloud data centers offer substantial computational power, with some of them directly attached to ground stations, while others are located at a significant distance from the ground station.
We denote this architecture as $G=(S \cup DC \cup GS \cup L)$, where $S$ denotes the set of low earth orbit satellites, $DC$ denotes the set of cloud data centers, $GS$ denotes the set of ground stations, and $L$ denotes the set of links among LEO satellites, ground station services, and cloud data centers, respectively.
% Define abbreviations and acronyms the first time they are used in the text, 
% even after they have been defined in the abstract. Abbreviations such as 
% IEEE, SI, MKS, CGS, ac, dc, and rms do not have to be defined. Do not use 
% abbreviations in the title or heads unless they are unavoidable.

\subsection{Inference Offloading and Partitioning}
Considering the characteristics of DNN models, task partitioning can be used to effectively alleviate the workload on satellites and minimize data transmission. 
DNN models typically  consist of various layers, including input, hidden, convolutional, pooling, and output layers, etc \cite{zhang2022comprehensive}, \cite{zhang2023comprehensive}. 
Each layer requires specific computing resources and incurs energy consumption. 
Taking advantage of this insight, a large-scale DNN task can be divided into subtasks corresponding to different layers. 
Notably, convolutional layers employ smaller kernels for feature extraction, while pooling layers reduce the spatial dimensions of feature maps through aggregation and downsampling.
Consequently, as the DNN network advances through its layers, the size of feature maps gradually decreases. 
Considering the limited computational payload of satellites, it is feasible to complete the DNN processing on the satellite up to a specified layer before offloading the remaining data and parameters to the ground.
This approach substantially reduces data transmission and processing on the satellite. 

To ensure the generalizability of our solution across different DNN models, we adopted the following settings. 
Given the diverse requirements and varying layer structures of DNN models for different tasks, we didn't concentrate on specific DNNs. 
Furthermore, numerous excellent experiments have already performed the partitioning of DNNs into subtasks based on layers \cite{b8}, \cite{b9}. 
Therefore, our focus is on the offloading process after partitioning a DNN model into distinct subtasks. 
The objective is to determine which layers of the DNN should be processed onboard the satellite and which layers should be offloaded for ground processing.
Accordingly, we denote $r$ as a request for a DNN model based on its layer structure. We divide $r$ into $\{M_{1},...M_{k},...M_{K}\}$, where $K$ is a positive integer and $1 \leq k \leq K$, $M_k$ represents the subtasks of request $r$. Let $D$ be the original data size of request $r${.}
% \begin{itemize}
% \item Use either SI (MKS) or CGS as primary units. (SI units are encouraged.) English units may be used as secondary units (in parentheses). An exception would be the use of English units as identifiers in trade, such as ``3.5-inch disk drive''.
% \item Avoid combining SI and CGS units, such as current in amperes and magnetic field in oersteds. This often leads to confusion because equations do not balance dimensionally. If you must use mixed units, clearly state the units for each quantity that you use in an equation.
% \item Do not mix complete spellings and abbreviations of units: ``Wb/m\textsuperscript{2}'' or ``webers per square meter'', not ``webers/m\textsuperscript{2}''. Spell out units when they appear in text: ``. . . a few henries'', not ``. . . a few H''.
% \item Use a zero before decimal points: ``0.25'', not ``.25''. Use ``cm\textsuperscript{3}'', not ``cc''.)
% \end{itemize}

\subsection{Latency}
An inference request for DNN models can be partitioned into $K$ tasks.
%, each requiring needs different computing resources and energy consumption. 
The overall latency is determined by the latency of each task. 
The latency can be divided into data processing and data transmission latency.

\textbf{Data processing latency.} The latency of a task is influenced by the computing resources available in the satellite and the cloud data center assigned to this task, as well as the data size being processed.
In line with our previously established settings, $D$ represents the original data size of request $r$. 
Additionally, the input matrix ratio of each layer is typically bounded and predetermined, denoted as $\alpha_{k}$. 
Consequently, the data size of each layer can be expressed as $\alpha_{k}\cdot{D}$. 
And the latency for processing $M_{k}$ in the satellite $S_i$ can be expressed as follows \cite{b11}, \cite{b12}, \cite{b13}:
\begin{equation}
\delta_{i,k} = {(\alpha_k\cdot D)}\cdot{\beta_i}\label{eq1}
\end{equation}
where $\beta_i$ is the latency for processing a unit amount of data of $S_i$, and the processing rate of different LEO satellites varies depending on their payloads and operating environment. Similarly, the latency for processing $M_{k}$ on the cloud data center can be expressed as follows:
\begin{equation}
\delta_{k}^{'} = {(\alpha_k\cdot D)}\cdot{\gamma}\label{eq2}
\end{equation}
where $\gamma$ represents the latency for processing a unit amount of data in the cloud data center.

\textbf{Data transmission latency.} Due to the long data transmission time, it is important to consider the latency associated with data transmission. When the satellite transmits data to the ground station, multiple factors come into play and need to be considered. Firstly, the satellite has a limited duration of contact with the ground station during each orbital period, typically around six minutes. Additionally, the connection between the satellite and the ground station undergoes dynamic changes, resulting in a relatively low transmission rate, usually less than 100 Mbps. Consequently, it is crucial to account for situations where the satellite cannot complete the transmission of all the required data within a single transmission cycle. Considering these factors, the latency of transmitting data from the satellite to the ground can be divided into two parts: the latency of satellite data transmission and the latency of waiting for data transmission when the ground station is out of contact with the satellite during the operational cycle. Accordingly, the latency for transmitting the input data of $M_{k}$ from the satellite to the ground can be expressed as follows:
\begin{equation}
t^{'}_{k}=t^{'}_{tr}+t^{'}_{per}=\frac{\alpha_k\cdot D}{R_{i}}+{t_{cyc}\cdot(\lceil \frac{\alpha_k\cdot D}{{R_{i}\cdot{t_{con}}}} \rceil-1)}\label{eq3}
\end{equation}
where $t^{'}_{tr}$ represents the transmission time for the input data of $M_{k}$, while $t^{'}_{per}$ denotes the waiting time during the transmission of $M_{k}$. $R_{i}$ corresponds to the transmission rate from Satellite $S_i$ to the ground station. Additionally, ${t_{cyc}}$ is the contact period time between the satellite and the ground station, and $t_{con}$ indicates the duration of communication between a satellite and a ground station.

In scenarios where the data is transmitted to a ground station without an associated cloud data center, the data must be further transmitted to a cloud data center located at a certain distance for processing.
In this case, the transmission delay of $M_{k}$ from ground station $p$ to cloud data center $q$ can be expressed as follows:
\begin{equation}
t_{g,c}=\frac{\alpha_k\cdot D}{R_{g_p,c_q}}\label{eq4}
\end{equation}
where ${R_{g_p,c_q}}$ represents the data rate through the channel from ground station $p$ to cloud data center $q$.

The overall task completion time of task $r$ comprises four components: the execution time on the satellite, the data transmission time from the satellite to the ground station, the data transmission time from the ground station to the cloud data center, and the subsequent task processing time in the cloud data center.
Let $h_k$ be a binary variable that indicates whether $M_{k}$ is executed on the satellite, where $h_k=1$ indicates that $M_{k}$ is executed on the satellite; otherwise $h_k=0$ indicates that $M_{k}$ is executed on the cloud data center. 
And $(h_{k-1}-h_k)$ indicates the layer from which data needs to be offloaded from the satellite to the ground.
Thus, the total latency of $r$ is given by:
\begin{equation}
\begin{split}
T=&T_{Satellite}+T_{StoG}+T_{GtoC}+T_{Cloud}\\
&=\sum_{k=1}^{{K}}{h_k}\cdot{\delta_{i,k}}+\sum_{k=1}^{{K}}{(h_{k-1}-h_{k})}\cdot{t^{'}_{k}}\\
&+\sum_{k=1}^{{K}}{(h_{k-1}-h_{k})}\cdot{t_{g,c}}+\sum_{k=1}^{{K}}{(1-h_k)}\cdot{\delta_{k}^{'}}\label{eq=5}
\end{split}
\end{equation}

% Number equations consecutively. To make your 
% equations more compact, you may use the solidus (~/~), the exp function, or  
% appropriate exponents. Italicize Roman symbols for quantities and variables, 
% but not Greek symbols. Use a long dash rather than a hyphen for a minus 
% sign. Punctuate equations with commas or periods when they are part of a 
% sentence, as in:
% \begin{equation}
% a+b=\gamma\label{eq}
% \end{equation}

% Be sure that the 
% symbols in your equation have been defined before or immediately following 
% the equation. Use ``\eqref{eq}'', not ``Eq.~\eqref{eq}'' or ``equation \eqref{eq}'', except at 注意如何引用的方法！
% the beginning of a sentence: ``Equation \eqref{eq} is . . .''

\subsection{Energy Consumption}
The total energy consumed by task $r$ is the sum of the energy consumed by each subtask. However, the satellite's computational power is constrained by its payload capacity, limited heat dissipation capabilities in space, and the low energy acquisition rate of solar panels. 
Consequently, the energy resources of satellites are severely restricted. Conversely, cloud data centers have an abundant electricity supply. 
Therefore, from a practical perspective, our primary focus is on conserving energy on the satellite. We further divide the energy consumption on the satellite into two components: during data processing and data transmission.

\textbf{Energy consumption during data processing.}  Previous research studies \cite{b14} and \cite{b15} have demonstrated that the energy consumption of a processing unit is directly proportional to both the task access rate and its maximum power consumption. 
%Furthermore, it is evident that the energy consumption of the computing payload primarily arises from the additional computational capabilities \cite{b16}. Notably, within the context of satellite operations, the energy consumption during computations primarily arises from the processing of DNN tasks. 
Therefore, our focus centers on the energy consumption model of DNN processing units. As a result, the energy consumption associated with executing $M_{k}$ of $r$ in $S_i$ can be expressed as follows:
\begin{equation}
e_{i,k}^{satellite}=\delta_{i,k}\cdot{(\frac{{\alpha_k\cdot D}}{\zeta_{i}\cdot\delta_{i,k}}{P_i^{max}}+P_i^{idel}+P_i^{leak})}\label{eq=6}
\end{equation}
where $\delta_{i,k}$ represents the latency for processing $M_{k}$ in $S_i$, $\zeta_{i}$ indicates the maximum amount of data that can be processed per unit time with maximum power, $P_i^{max}$ denotes the maximum power consumption of all GPU units on satellite $i$, $P_i^{idle}$ represents the idle power consumption of satellite $i$, and $P_i^{leak}$ refers to the power leakage of the GPU units on satellite $i$.

\textbf{Energy consumption during data transmission.} Recall that $t^{'}_{tr}$ denotes the transmission time for the input data of $M_{k}$, while $P_i^{off}$ represents the data transmission power of the antenna satellite $i$. The energy consumption of satellite $i$ during the transmission of the input data of $M_{k}$ (or the output data of $M_{k-1}$) to the ground can be expressed as:
\begin{equation}
e_{i,k}^{off}=t^{'}_{tr}\cdot{P_i^{off}}=\frac{\alpha_k\cdot D}{R_{i}}\cdot{P_i^{off}}\label{eq=7}
\end{equation}

Based on the aforementioned analysis, the comprehensive energy consumption during the execution of a DNN inference task can be formulated as follows:
\begin{equation}
E=\sum_{k=1}^{{K}}{h_k}\cdot{e_{i,k}^{satellite}}+\sum_{k=1}^{{K}}{(h_{k-1}-h_k)}\cdot{e_{i,k}^{off}}\label{eq=8}
\end{equation}

\subsection{Problem Formulation}
The objective of this model is to effectively address the optimization problem of minimizing energy consumption and reducing task latency. This is achieved through the implementation of efficient partitioning and offloading techniques for DNN models throughout the entire DNN inference process. 
Considering the diverse nature of satellite tasks, certain specific tasks require meeting low-latency requirements, especially critical applications like fire hazard detection.
Conversely, for longer-duration detection tasks, 
such as remote sensing observations of changes in terrain and landforms, conserving valuable energy resources on the satellite becomes of utmost importance.
Considering the varying importance of energy and latency in different applications, we incorporate both factors into the optimization objective by applying appropriate weights.
Furthermore, to account for the disparate magnitudes of energy and latency and to ensure that both factors are effectively reflected in the objective function, normalization is applied to both energy and latency.
Consequently, the problem can be formulated as an integer linear problem, expressed as follows:
\begin{equation}
min~~Z=\mu\frac{E-E_{min}}{E_{max}-E_{min}}+\lambda\frac{T-T_{min}}{T_{max}-T_{min}}\label{eq=9}
\end{equation}

s.t.:

% \begin{equation}
% \beta_i\geq 
% \left\{
% \begin{aligned}
%     &{\beta_{max}}~~In~the~sunlit.\\
%     &{\beta_{max}^{'}}~~In~the~dark.
% \end{aligned}
% \right.
% \label{eq=10}
% \end{equation}

% \begin{equation}
% {\mu+\lambda=1}\label{eq=17}
% \end{equation}

\begin{equation}
{\gamma}\geq {\gamma_{max}}\label{eq=11}
\end{equation}

% \begin{equation}
% T\leq{T_{max}}\label{eq=12}
% \end{equation}

\begin{equation}
\sum_{k=1}^{{K}}{h_k}+\sum_{k=1}^{{K}}({1-h_k})=K\label{eq=13}
\end{equation}

\begin{equation}
\sum_{k=1}^{{K}}{(h_{k}-h_1)}\leq{1}\label{eq=14}
\end{equation}

\begin{equation}
{h_k}\geq{h_{k+1}}\label{eq=15}
\end{equation}
  
\begin{equation}
h_k\in {\{0,1\}}\label{eq=16}
\end{equation} 
where $\mu$ and $\lambda$ are weighting coefficients, ${\mu+\lambda=1}$. And $Z$ is the optimization objective.
% Eq. (\ref{eq=17}) ensures that the sum of the weights assigned to energy consumption and time is equal to 1, enabling a clear understanding of the relative importance of the two factors within the overall context. 
Eq. (\ref{eq=11}) specifies the upper limit on the latency for processing a unit amount of data in a cloud data center. Eq. (\ref{eq=13}) guarantees the integrity of request $r$ by ensuring that all subtasks $M_{k}$ are executed either on the satellite or in the cloud data center. Eq. (\ref{eq=14}) indicates that when intermediate results of satellite execution need to be transmitted, there should be exactly one. Eq. (\ref{eq=15}) makes sure that the data before the down transmission is running on the satellite, and the data after the down transmission is running on the ground. Eq. (\ref{eq=14}) and Eq. (\ref{eq=15}) maintain the continuity of request $r$. Eq. (\ref{eq=16}) restricts the binary variables.
%Our objective is to minimize the overall result by determining the optimal strategy for task offloading. 
%Since both $E$ and $T$ can be represented using binary variables $h_k$, the problem is defined as an integer linear programming problem with constraints. 

The problem is an integer linear programming problem with integer constraints, resulting in a combinatorial explosion in the discrete solution space. So we need an algorithm to reduce complexity while obtaining an approximate optimal solution.
\begin{algorithm}[htb]
    \caption{Inference Offloading Algorithm}
    \label{alg:algorithm3}
    \LinesNumbered
        \KwIn {$G=(S \cup DC \cup GS \cup L)$, a sequence of tasks $M_{1},...M_{k},...,M_{K}$ that belong to $r$, which need to operation in the limited time. The weighting coefficients: $\lambda$ and $\mu$} %%input
        \KwOut {An offloading decision {$h_k$} for the inference request.}    %%output
        $H=\{h_1,...h_k,...h_K\}=\{0\}$\;\tcp{Initial offloading state}
        $Cons \leftarrow Constrains(10)-(14)$\;
        $Ans \leftarrow {inf}$\;
        ILPB({$H, Cons, Ans$})\;\tcp{Call ILPB function}
       Return best offloading decision $H$\;   
        \SetKwFunction{ILPB}{ILPB}
        \SetKwProg{Fn}{Function}{:}{}
        \Fn{\ILPB{{$\{h_k\}, Cons, Ans$}}}{
            \If{$|Ans'-Ans| < 1e-5$}
            {return;}\tcp{Recursive Termination Condition}
            \eIf{all $Cons$ satisfied}
            {\If{$Z<Ans$}
            {$Ans \leftarrow Z$\;\tcp{Update the optimal solution}
            update $H$\;\tcp{Update the offloading solution}
            }}
            {return;}
             \ForEach{undetermined $h_k$ \textbf{in} $H$}
             {\ForEach{possible value of $h_k$}
             {\If{$Z(h_k)+minZ({\{\overline{h_k}\}}) < Ans$}
             {update $H$\;
             $Ans \leftarrow Z(h_k)+minZ({\{\overline{h_k}\}})$\;
             $Ans'$ = ILPB({$H, Z_{\text{Min}}, Cons, Ans$})\;\tcp{Recursively call the ILPB function to the next state}}
             
        }}}
\end{algorithm}
\section{Inference Offloading Algorithm}
We exploit the branch and bound technique to tackle the integer linear programming problem.
The branch and bound method continuously partitions the problem space and applies bounding conditions to each subproblem to discovery the optimal solution or  approximation optimal solution. 
Despite being an exhaustive search algorithm, it efficiently reduces the search space by intelligently pruning unnecessary branches.
The key idea involves expanding the feasible solution space of the problem as a branching tree and searching for the optimal solution within each branch.

The detailed algorithm is given in Algorithm 1, which is referred to as integer linear programming based on branch and bound method (ILPB). 
First, initialize the decision variables $H$, and set the upper bound of the optimal solution (lines 1-3). 
Next, call the ILPB function and return the offloading optimal decision (lines 4-5).
The remaining part is the definition of the ILPB function.
When the recursion boundary is reached, return the result (lines 7-9).
If the current $H$ satisfies the constraint conditions and is a feasible solution, evaluate the solution. 
If the current solution $Z$ is better than the optimal solution $Ans$, update the optimal solution to the current solution; otherwise, Terminate this condition and return. (lines 10-17). 
Select an undecided integer variable $h_k$, and for each possible value of the variable, if the current function value plus the minimum possible value of the remaining variables is less than the optimal solution $Ans$, update the set of decision variables and the optimal solution.
Call the ILPB function in the current situation (lines 18-26). 
This approach effectively reduces the computational complexity by eliminating the consideration of numerous subsets. The presence of integer constraints in the problem results in a combinatorial explosion within the search space.
% In the context of our minimizing problem, the entire feasible solution space is partitioned into smaller subsets through iterative branching operations. Additionally, a bounding operation is implemented to compute a lower bound for the solution set within each subset. Subsets whose bounds exceed the target value of the known feasible solution set are no longer further branched after each branching step. This approach effectively reduces the computational complexity by eliminating the consideration of numerous subsets. The presence of integer constraints in the problem results in a combinatorial explosion within the search space.

% The detailed algorithm is given in Algorithm 1, which is referred to as integer linear programming based on branch and bound method(ILPB).

\section{EXPERIMENTS}
In this section, we evaluate the performance of our proposed algorithm by comparing it with the following two algorithms:
\begin{itemize}
\item \textbf{All tasks are offloaded to the ground (ARG):} 
The satellite transmits all captured data to the ground station, which then transfers them to the cloud data center for centralized processing.
% Similar to the architecture of the “bent pipe”, the satellite solely performs data forwarding and does not engage in data processing tasks.
\item \textbf{All tasks are completed on the satellite (ARS):} 
The satellite autonomously executes the entire DNN task on its onboard payload.
%Since the data volume of the final results is typically extremely small, the latency and energy consumption incurred by transmitting the final results to the ground station can be neglected. 
% Only the total energy consumption and total latency generated by the satellite during the execution process are calculated.
\end{itemize}
\begin{figure*}[t] % 使用figure*环境实现双栏宽度的图片
  \centering
  \begin{minipage}[b]{0.32\linewidth}
    \includegraphics[width=\linewidth]{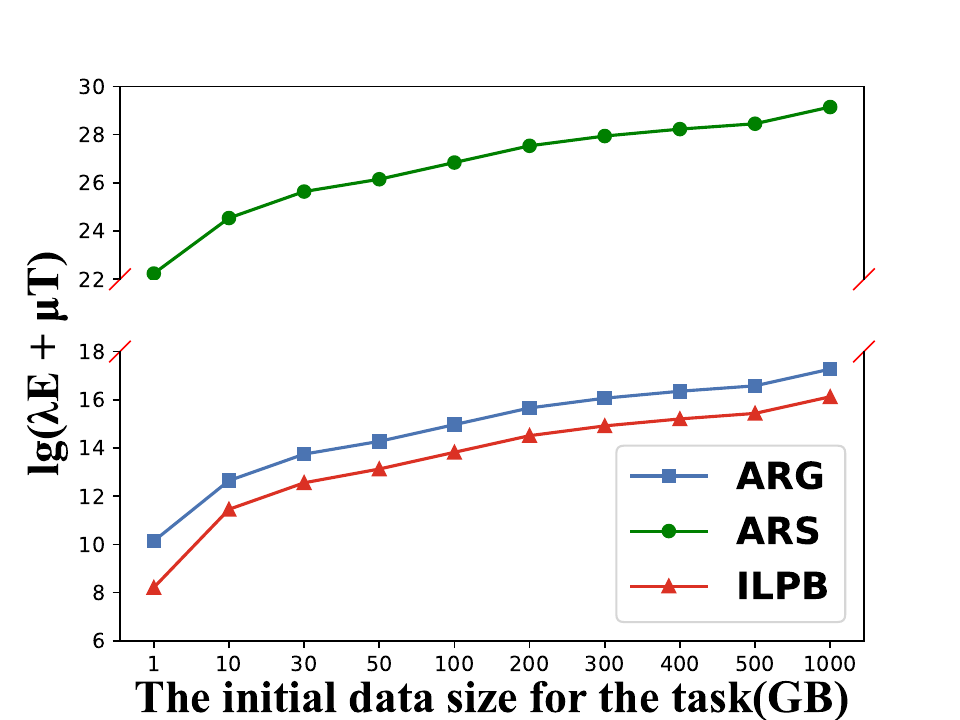}
    \caption{The total consumption of tasks in different initial data size.}
    \label{fig2}
  \end{minipage}
  \hfill
  \begin{minipage}[b]{0.32\linewidth}
    \includegraphics[width=\linewidth]{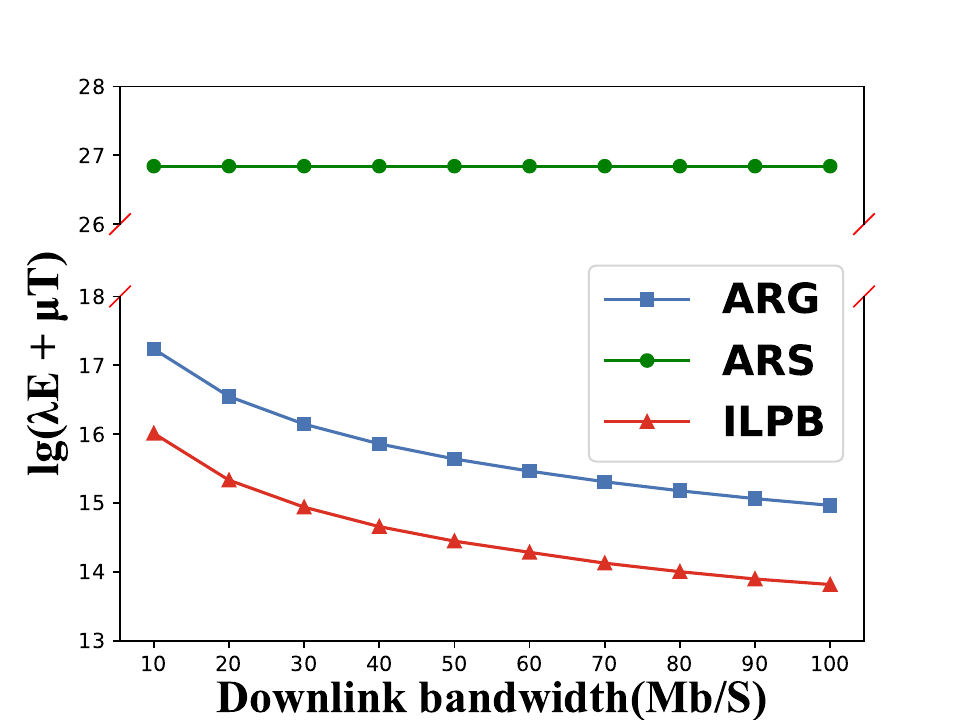}
    \caption{The total consumption of tasks in different transmission rate.}
    \label{fig3}
  \end{minipage}
  \hfill
  \begin{minipage}[b]{0.32\linewidth}
    \includegraphics[width=\linewidth]{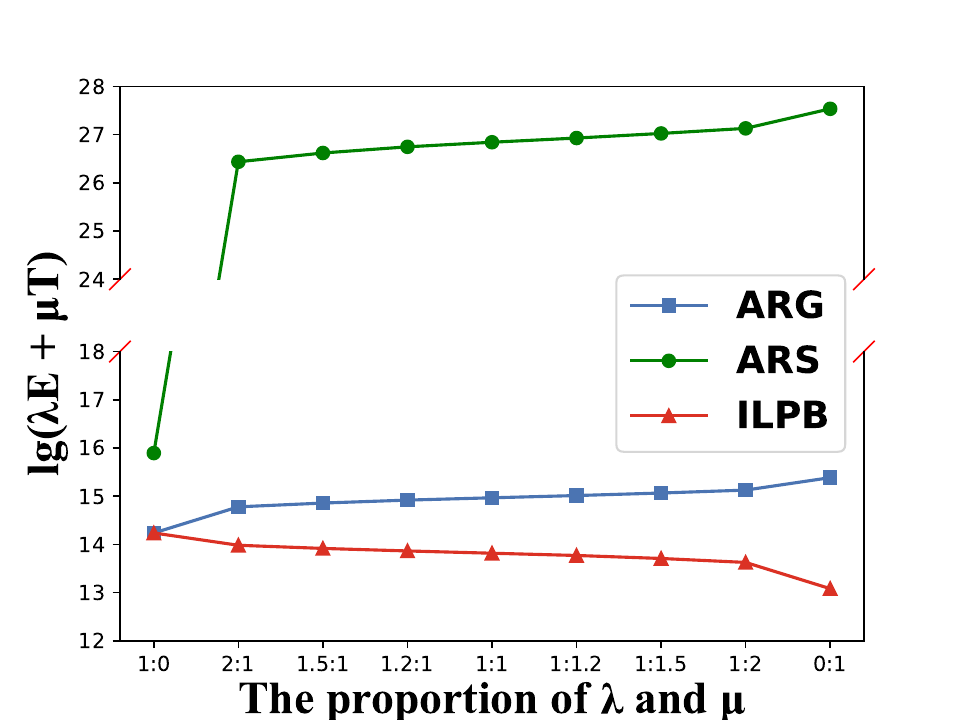}
    \caption{The total consumption of tasks in different portion for time and energy.}
    \label{fig4}
  \end{minipage}
\end{figure*}
% \begin{figure}[t]
% % \centerline{\includegraphics{fig.png}}
% \centering
% \includegraphics[width=0.5\textwidth]{Figure_dataSize.png}
% \caption{The total consumption of tasks in different initial data size.
% }
% \label{fig2}
% \end{figure}

% \begin{figure}[t]
% % \centerline{\includegraphics{fig.png}}
% \centering
% \includegraphics[width=0.5\textwidth]{Figure_transRatepng.png}
% \caption{The total consumption of tasks in different transmission rate between the satellite and the ground.
% }
% \label{fig3}
% \end{figure}
\newcommand{\upcite}[1]{\textsuperscript{\textsuperscript{\cite{#1}}}}
\subsection{Experiment Setup}
In the simulation, we utilize the experiment parameters from a real-world LEO satellite constellation named “Tiansuan~Constellation”\footnote{http://www.tiansuan.org.cn/}. 
Satellites of this constellation operate at an orbital altitude of approximately 500 km and pass over a ground station for data transmission every 8 hours. 
The data transmission duration per pass is approximately 6 minutes, and the transmission rate fluctuates within the range of [10, 100] Mbps. 
The parameters $\beta_i$ and $\gamma_j$ on processing unit amount (1 KB) of data in the satellite and cloud data center are varied from ranges [0.01, 0.03] seconds and [0.0001, 0.001] seconds, respectively. 
The parameter $\alpha_j$ is varied from ${0.05}^k$ to ${0.9}^k$, considering the reduction in the size of the input data for a layer compared to the input data of the previous layer. The max processing powers of a satellite are set to [1, 10] Watt\cite{b16}. 
And the size of input computation task is set to [1,1000] GB. 
Due to the large values of energy and time consumption, we represent the results after applying a logarithmic transformation.
\subsection{Experiment Results}
As illustrated in Fig.~\ref{fig2}, the energy and time consumption of the three algorithms varies with different initial data sizes. 
As expected, all three algorithms demonstrate increased energy and time consumption with a larger initial data size. 
The ILPB algorithm consistently outperforms others in terms of energy and time consumption. 
Moreover, it exhibits a slower growth rate as the initial data size increases.
Our method achieves a significant reduction in overall time and energy consumption, amounting to just 10\%-18\% of the average values obtained from ARG plus ARS.

Next, we study the energy and time consumption associated with the IPLB, ARG, and ARS algorithms under varying data transmission rates between the satellite and the ground station.
As illustrated in Fig.~\ref{fig3}, the transmission rate is ranged from 10 to 100 MB/s with a step size of 10. 
The results demonstrate that the proposed IPLB algorithm consistently achieves lower energy consumption and latency compared to the ARG and ARS algorithms. 
Moreover, as the data transmission rate between the satellite and the ground increases, both the IPLB and ARG demonstrate decreased total time and energy consumption required on the satellite.
However, the energy consumption of the ARS method remains largely unaffected by the data transmission rate due to its exclusive execution of tasks on the satellite.

Finally, we study the energy and time consumption of the IPLB, ARG, and ARS algorithms across various proportions of time and energy, as illustrated in Fig.~\ref{fig4}. 
When the ratio between $\lambda$ and $\mu$ is set to 1:0, indicating that energy consumption is not considered, both the ILPB and ARG algorithms demonstrate comparable total time delays for task execution. 
Notably, the total time of both the ILPB and ARG algorithms is lower than that of the ARS algorithm.
However, when the ratio between $\lambda$ and $\mu$ is set to 0:1, the ILPB algorithm surpasses the ARG algorithm by a substantial margin, indicating its superiority in energy consumption. 
Moreover, as the proportion of $\mu$ increases, the ILPB algorithm significantly reduces energy consumption while maintaining shorter time delays, thereby conserving valuable energy resources on the satellite.

% \begin{figure}[htbp]
% % \centerline{\includegraphics{fig.png}}
% \centering
% \includegraphics[width=0.5\textwidth]{Figure_ratio.png}
% \caption{XXXXXXX
% }
% \label{fig4}
% \end{figure}

% \begin{figure}[t]
% % \centerline{\includegraphics{fig.png}}
% \centering
% \includegraphics[width=0.5\textwidth]{Figure_proportion.png}
% \caption{The total consumption of tasks in different portion for time and energy.
% }
% \label{fig4}
% \end{figure}

\section{CONCLUSION}
This paper addresses the inference offloading problem via satellite-ground collaborative computing, aiming to address challenges associated with limited energy acquisition, insufficient computational resources, and poor satellite-ground communication. 
We utilize the hierarchical structure of DNNs to partition tasks into subtasks and establish an inference offloading scheme based on a comprehensive metric that encompasses both delay and energy consumption. 
We incorporate the branch and bound method into the ILP framework to achieve efficient problem-solving.
Simulation results validate the effectiveness of the proposed ILPB algorithm in reducing both time and delay. 
Under different initial data sizes and transmission rates, the algorithm proposed in this paper outperforms the other two strategies.
We believe that our approach delivers valuable information and provides an excellent tool that facilitates a more detailed exploration of the inference tasks of LEO satellites.
The future work can reduce task complexity through some model lightweight techniques.
%Our research focuse exclusively on the task offloading problem between a single satellite and the ground.
%Future research will investigate task offloading for coordinated computing among multiple satellites and the ground.
\section{Acknowledgement}
This work was supported by the National Natural Science Foundation of China under grants U21B2016 and 62032003.
\bibliographystyle{IEEEtran}
\bibliography{ref}

% \begin{thebibliography}{00}
% % \bibitem{b1} G. Eason, B. Noble, and I. N. Sneddon, ``On certain integrals of Lipschitz-Hankel type involving products of Bessel functions,'' Phil. Trans. Roy. Soc. London, vol. A247, pp. 529--551, April 1955.
% % I Del Portillo, BG. Cameron, and EF. Crawley. ``A technical comparison of three low earth orbit satellite constellation systems to provide global broadband,'' Acta astronautica, 2019, 159 123-135.
% \bibitem{b2} J. Clerk Maxwell, A Treatise on Electricity and Magnetism, 3rd ed., vol. 2. Oxford: Clarendon, 1892, pp.68--73.
% \bibitem{b3} I. S. Jacobs and C. P. Bean, ``Fine particles, thin films and exchange anisotropy,'' in Magnetism, vol. III, G. T. Rado and H. Suhl, Eds. New York: Academic, 1963, pp. 271--350.
% \bibitem{b4} K. Elissa, ``Title of paper if known,'' unpublished.
% \bibitem{b5} R. Nicole, ``Title of paper with only first word capitalized,'' J. Name Stand. Abbrev., in press.
% \bibitem{b6} Y. Yorozu, M. Hirano, K. Oka, and Y. Tagawa, ``Electron spectroscopy studies on magneto-optical media and plastic substrate interface,'' IEEE Transl. J. Magn. Japan, vol. 2, pp. 740--741, August 1987 [Digests 9th Annual Conf. Magnetics Japan, p. 301, 1982].
% \bibitem{b7} M. Young, The Technical Writer's Handbook. Mill Valley, CA: University Science, 1989.
% \end{thebibliography}
\vspace{12pt}
% \color{red}
% IEEE conference templates contain guidance text for composing and formatting conference papers. Please ensure that all template text is removed from your conference paper prior to submission to the conference. Failure to remove the template text from your paper may result in your paper not being published.

\end{document}